\documentclass[twocolumn]{aastex631}
\usepackage{bm}
\usepackage{epsfig}
\usepackage[utf8]{inputenc}
\usepackage{ulem}
\usepackage{graphicx}
\usepackage{color}
\usepackage{subfigure}
\newcommand{\be}{\begin{equation}}
\newcommand{\ee}{\end{equation}}
\newcommand{\beq}{\begin{equation}}
\newcommand{\eeq}{\end{equation}}
\newcommand\bea{\begin{eqnarray}}
\newcommand\eea{\end{eqnarray}}
\renewcommand{\d}{\partial}

\shorttitle{turbulence and particle acceleration}
\shortauthors{Vega et al.}

\begin{document}
\title{Turbulence and particle acceleration in a relativistic plasma}

\correspondingauthor{Cristian Vega}
\email{csvega@wisc.edu}

\author{Cristian Vega}
\affiliation{Department of Physics, University of Wisconsin at Madison, Madison, Wisconsin 53706, USA}

\author[0000-0001-6252-5169]{Stanislav Boldyrev}
\affiliation{Department of Physics, University of Wisconsin at Madison, Madison, Wisconsin 53706, USA}
\affiliation{Center for Space Plasma Physics, Space Science Institute, Boulder, Colorado 80301, USA}

\author[0000-0003-1745-7587]{Vadim Roytershteyn}
\affiliation{Center for Space Plasma Physics, Space Science Institute, Boulder, Colorado 80301, USA}

\author{Mikhail Medvedev}
\affiliation{Department of Physics and Astronomy, University of Kansas, Lawrence, KS 66045, USA}
\affiliation{Laboratory for Nuclear Science, Massachusetts Institute of Technology, Cambridge, MA 02139, USA}

\date{\today}

\begin{abstract}
 In a collisionless plasma, the energy distribution function of plasma particles can be strongly affected by turbulence. In particular, it can develop  a non-thermal power-law tail at high energies. We argue that turbulence with initially relativistically strong magnetic perturbations (magnetization parameter $\sigma \gg 1$) quickly evolves into a state with ultra-relativistic plasma temperature but mildly relativistic turbulent fluctuations.  We present a phenomenological and numerical study suggesting that in this case, the exponent $\alpha$ in the power-law particle energy distribution function, $f(\gamma)d\gamma\propto \gamma^{-\alpha}d\gamma$, depends on magnetic compressibility of turbulence. Our analytic prediction for the scaling exponent $\alpha$ is in good agreement with the numerical results.
\end{abstract}

\keywords{}




\section{Introduction}
Large-scale astrophysical flows are often unstable, which leads to the generation of strongly interacting fluctuations covering a broad range of spatial and temporal scales. The resulting turbulence is essentially a non-equilibrium phenomenon where relatively slowly evolving large-scale modes contain most of the fluctuation energy. Due to nonlinear interactions, the energy is transferred to smaller scales and eventually gets removed from the collective plasma motion by various ``non-ideal'' mechanisms, such as kinetic viscosity, resistivity, or radiative processes.  

The situation becomes nontrivial when a turbulent plasma is collisionless, a condition rather well satisfied in many astrophysical environments \cite[e.g.,][]{elmegreen2004,chen2016}. This happens not only because the mechanisms of energy conversion from the collective motion to particle kinetic energy become more complex, but also because particles interacting with turbulent fluctuations, may not have enough time for collisional relaxation and as a result end up forming non-Maxwellian tails in the distribution function. 

Recently, pioneering numerical studies have been conducted of relativistic collisionless plasma turbulence  \cite[][]{zhdankin2017a,zhdankin2018,zhdankin2018c,zhdankin2019,zhdankin2021,comisso2018,comisso2019,comisso2021,zhdankin2020,nattila2020}, which may be relevant for pulsar wind nebulae, coronae of accretion disks around black holes, active galactic nuclei jets, and other high-energy astrophysical phenomena. Such turbulence is also interesting on its own as a relatively little explored regime of nonlinear plasma dynamics.  These studies have been performed for various settings of turbulence: decaying, driven, radiative. They have discovered that the energy distribution functions of plasma particles in such turbulence universally develop non-thermal power-law tails, $f(\gamma)\,d\gamma \propto \gamma^{-\alpha}\,d \gamma$, at ultra-relativistic energies $\gamma\gg 1$, with the scaling exponents satisfying $\alpha \lesssim 3$. 

In this work, we study phenomenologically and numerically the features of plasma turbulence which is driven by strong initial magnetic perturbations such that their energy exceeds the rest-mass energy of plasma particles. We argue that such turbulence rapidly evolves to a state where plasma motion is mildly relativistic while plasma temperature is ultra-relativistic. We propose that locally, a fraction of particles may be trapped by spontaneously generated magnetic structures where the particles get accelerated. Due to pitch-angle scattering, the particles eventually leave the trapping regions.  The pitch-angle width of the trapped population depends on the so-called magnetic compressibility of turbulence, and it defines the power-law exponent of the resulting particle energy distribution function. Below we first present a phenomenological discussion of relativistic plasma turbulence and a model for associated particle acceleration, and then compare our predictions with the 2D particle-in-cell numerical simulations of pair plasma.

\section{Relativistic turbulence in a plasma}
As in the non-relativistic case, useful insight into relativistic plasma dynamics is given by the analysis of conservation laws. For simplicity, let us first assume that a plasma is relativistically hot and that turbulence consists of fluid elements moving with ultra-relativistic velocities. The energy of a fluid element of volume $\Delta V$ (consisting of a given number of particles) is then \cite[e.g.,][]{mihalas1984}
\begin{eqnarray}
{\cal E}=\left\{{\tilde \gamma}^2\left[\rho_0 c^2+\rho_0\epsilon_0+p \right]-p\right\}\Delta V = \nonumber \\
m \left\{ {\tilde \gamma}\left[c^2+\epsilon_0+p/\rho_0 \right]-p/{(\rho_0\tilde \gamma})\right\}\approx \nonumber \\
m{\tilde \gamma}\left[c^2 +\epsilon_0+p_0 \right],
\end{eqnarray}
and its momentum is given by
\begin{eqnarray}
{\bf P}= m{\tilde \gamma}\left[c^2 + \epsilon_0+p_0 \right]\frac{\bf v}{c^2}.
\end{eqnarray}
Here, ${\bf v}$ is the plasma-fluid velocity and ${\tilde \gamma}\gg 1$ is the Lorentz factor of fluid motion (we use the tilde sign to distinguish it from the gamma factor of particles). In this equation, $\rho_0$ is the mass density, $\epsilon_0$ is the specific internal energy, $p$ is plasma pressure, all measured in the fluid element frame, and $p_0\equiv p/\rho_0$. (In our qualitative discussion we assume that the particle distribution is isotropic in that frame.)  We also used the fact that for a given number of particles, the mass of a fluid element, $m={\tilde \gamma}\rho_0 \Delta V$ is constant. To simplify the notation even further, we incorporate the internal energy and pressure into the definition of the effective mass of a fluid element ${\tilde m}=m[1+\epsilon_0/c^2+p_0/c^2]$, so that we have familiar-looking expressions for the energy and momentum, ${\cal E}={\tilde \gamma}{\tilde m}c^2$ and ${\bf P}={\tilde \gamma}{\tilde m}{\bf v}$. 

Consider now two fluid elements colliding with each other, and assume that such a collision is inelastic, so that a new fluid element of effective mass ${\tilde M}$ is formed. It then follows that after a typical collision, the effective mass of the new element is significantly larger than the effective masses of the original elements. To show that, we use the energy and momentum conservation laws to write 
\begin{eqnarray}\label{energy}
\sqrt{{\tilde M}^2c^4+\left({\bf P}_1+{\bf P}_2 \right)^2c^2}= P_1 c+P_2 c,
\end{eqnarray}
from which we get 
\begin{eqnarray}\label{mass}
{\tilde M}=\frac{\sqrt{2P_1P_2}}{c}\sqrt{1-\cos\phi}\sim {\tilde m}{{\tilde \gamma}},
\end{eqnarray}
where we assumed that the colliding elements have similar masses and energies. Here, $\phi$ is the angle between the initial momenta. Since ${\tilde \gamma}\gg 1$, the internal energy of fluid elements increases exponentially fast with the number of such ultra-relativistic collisions. 
Simultaneously, 
one can derive from Eqs.~(\ref{energy}), (\ref{mass}) for the final gamma factor
\begin{eqnarray}
{\tilde \gamma}_f=\frac{P_1+P_2}{\sqrt{2P_1P_2}}\frac{1}{\sqrt{1-\cos\phi}},
\end{eqnarray}
so that ${\tilde \gamma}_f \gtrsim 1$ if ${\tilde \gamma}_1\sim {\tilde \gamma}_2$.  Therefore, the gamma factors of newly formed fluid elements decrease fast with the number of collisions. 
It is therefore reasonable to propose that initial ultrarelativistic turbulent motion rapidly evolves into mildly relativistic (${\tilde \gamma}-1\ll 1$) turbulence of ultra-relativistically hot ($\gamma\gg 1$) plasma. Such a statement seems to be consistent with available simulations of relativistic plasma turbulence \cite[][]{zhdankin2018c,comisso2019}, and with the results of our numerical studies presented below.

When electromagnetic field is taken into account, the energy consists of the magnetic, electric, and kinetic contributions.  The energy density is
\begin{eqnarray}
w=\frac{B^2}{8\pi}+\frac{E^2}{8\pi}+\tilde{\gamma}^2\left(\epsilon +p\frac{v^2}{c^2}\right),
\end{eqnarray}
where $\epsilon =\rho_0c^2+\rho_0\epsilon_0$ is the plasma particle energy density calculated in the rest frame of fluid element.  Let us assume ultrarelativistic plasma temperature, so that we have for the equation of state $p\approx \epsilon/3$. For non-relativistic  plasma fluctuations, the kinetic energy is approximated as
$\tilde{\gamma}^2\left(\epsilon +p\frac{v^2}{c^2}\right) \approx \epsilon +\frac{4\epsilon}{3}\frac{v^2}{c^2}.$
Restricting ourselves to the energy associated with turbulent fluctuations,\footnote{Here we assume that as plasma is heated by turbulence, it mostly increases the uniform component of~$\epsilon$ rather than its fluctuations. For instance, numerical simulations of \cite[][]{zhdankin2018c} show that as turbulence evolves, the internal energy continuously increases with time and significantly exceeds the magnetic and bulk kinetic energies. The spectrum of the internal energy at $k>k_0$ (where $k_0$ corresponds to the outer scale of turbulence), however, remains significantly below that of the magnetic and kinetic energies. Therefore, as plasma is heated by turbulence in this case, most of the internal energy is accumulated at low-$k$ modes.}  we then have
\begin{eqnarray}
\label{delta_w}
\delta w=\frac{(\delta B)^2}{8\pi}+\frac{E^2}{8\pi}+\frac{4\epsilon }{3} \frac{v^2}{c^2}.
\end{eqnarray}

Assuming that the plasma is a good conductor, ${\bf E}\approx -{\bf v}\times {\bf B}_0/c$, we observe that the ratio of the electric energy term to the kinetic one is essentially the magnetization parameter of the plasma, that is,
\begin{eqnarray}
\frac{E^2}{\epsilon v^2/c^2}\sim \frac{B^2_0}{\epsilon}= \sigma.
\end{eqnarray}
When the magnetization is weak, $\sigma\ll 1$, the electric fluctuations are negligible, and the turbulent energy is given by the magnetic and kinetic terms. In the case of strong magnetization, $\sigma\gg 1$, the energy is mostly contained in electric and magnetic fluctuations. 

In both relativistic and non-relativistic Alfv\'enic turbulence, it is reasonable to assume equipartition of energy among fluctuating fields. This holds for the linear Alfv\'en waves \cite[e.g.,][]{thompson1998,keppens2008,lemoine2016,demidem2020,tenbarge2021,mallet2021}, but is also satisfied approximately in a nonlinear turbulent regime, which is the subject or our study. We see from Eq.~(\ref{delta_w}) that in the case $\sigma\ll 1$, an argument that the magnetic and kinetic fluctuations are on the same order leads to an estimate $v/v_A\sim \delta B/B_0$, where we use the definition of the Alfv\'en speed: $v_A=B_0/\sqrt{4\pi(\epsilon+p)}$. In the opposite case of strong magnetization, $\sigma \gg 1$, the equipartition of magnetic and electric contributions gives $v/c\sim \delta B/B_0$. In this case, when the fluid motion is non-relativistic, the magnetic fluctuations are small as well. 

In order to understand why electric energy becomes significant in the case of large $\sigma$, consider an analogy with a {\it non-relativistic} plasma. For that, let us see what happens when the internal energy becomes comparable to the rest-mass energy of a fluid element, that is, $\epsilon \sim n mc^2$. As one can check, the magnetization parameter then turns into the so-called quasineutrality parameter, $\sigma = \Omega_i^2/\omega_{pi}^2$.  Here, $\Omega_i$ and $\omega_{pi}$ are, correspondingly, the gyrofrequency and plasma frequency of the ions if a plasma is composed of electrons and ions (a similar consideration can be conducted for a pair plasma). 
In most laboratory and natural applications, the quasineutrality parameter is small, $\sigma \ll 1$, in which case the plasma is quasi-neutral so that the fluctuations of the electric charge can be neglected. As a result, the electric energy is negligible as compared to the kinetic energy of plasma fluctuations. However, in the opposite limit, $\sigma \gg 1$, the charge fluctuations cannot be neglected and the electric energy dominates.

\section{Particle energization} 
Collisional dissipation is not the only mechanism by which energy is transferred from collective turbulent fluctuations to particles. Particle heating in a weakly collisional plasma may also be mediated by particle interactions with turbulent fluctuations; we will call such processes ``anomalous" heating. It has been observed in recent kinetic numerical studies \cite[e.g.,][]{zhdankin2017a,comisso2018,zhdankin2018c,zhdankin2019,zhdankin2020,comisso2019,nattila2020,zhdankin2021} that in a relativistically hot turbulent plasma, the particle energy distribution function develops a non-thermal tail at high energies, approximately described by a power law.  Analytical description of anomalous heating is a difficult task. We will therefore base our discussion on the two key numerical observations related to the acceleration process. 

First, the particle energization resembles a diffusion process in the energy space, with the diffusion coefficient scaling with the particle energy as $ \gamma^2$ at $\gamma\gg 1$.   This may indicate that the particles experience a second-order Fermi acceleration \cite[e.g.,][]{fermi1949,teller1954,kulsrud1971}, where scattering events are provided by moving magnetic mirrors (e.g., turbulent eddies or magnetic structures advected by turbulence \cite[][]{achterberg1984,selkowitz2004,yan2008,peera2014,demidem2020}). An individual scattering event then randomly changes the particle energy as $\Delta \gamma\propto \pm\gamma$, see, e.g., the discussion in \cite[][]{zhdankin2018c,blandford1987}. It is also consistent with particle interactions with wave turbulence, since as was elucidated in \cite[][]{demidem2020}, the quasilinear diffusion coefficient of ultrarelativistic particles in strong Alfv\'enic turbulence is also proportional to $\gamma^2$.  In addition, numerical analysis of \cite[][]{comisso2019} suggests that eventual diffusive energization by turbulent eddies is preceded by initial fast acceleration by reconnection events. Second, numerical simulations indicate the presence of an exponentially strong regular dissipation process 
at large energies \cite[e.g., Figure 4 in][]{zhdankin2020}, which is crucial for establishing a power-law distribution of ultra-relativistic particles.


We propose that a power-law particle energy distribution may be understood based on the following phenomenological dynamical model that incorporates both observed diffusion and dissipation. Consider an ultrarelativistic particle with momentum $p\equiv |{\bf p}|\approx p_0$, and describe particle interactions with non-relativistic randomly moving fluid elements by a stochastic dynamical equation (Langevin equation):
\begin{eqnarray}
d {\bf p}/dt=p\,\bm{\eta}(t),
\end{eqnarray}
with an isotropic Gaussian white random noise
\begin{eqnarray}
\label{eta}
\langle \eta^i(t)\eta^j(t') \rangle=2D\delta^{ij}\delta(t-t'),\\
\langle \eta^i(t)\rangle=0,
\end{eqnarray}
where $D$ is a constant normalization coefficient. Obviously, such an equation satisfies the required diffusion scaling $\langle (\Delta p)^2\rangle \propto p^2$, while the random vector $\bm{\eta}$ mimics the velocity of the scatterers. The corresponding Fokker-Planck equation for the probability density function $F({\bf p})$, can then be easily derived~\cite[e.g,][]{oksendal2003}:
\begin{eqnarray}
\frac{\partial F}{\partial t}=D\frac{\partial }{\partial p^i} p \frac{\partial }{\partial p^i}\left( p F\right),
\label{fokker_planck}
\end{eqnarray}
where we sum over repeated indices. As there is a local mean magnetic field in a region where particles get accelerated,  we may re-write Eq.~(\ref{fokker_planck}) in the spherical coordinates with respect to the direction of the field. Introducing the phase-space-volume compensated function $f(p,\mu)=F(p,\mu)4\pi p^2$, we get
\begin{eqnarray}
\label{fp_2}
\frac{\partial f(p, \mu)}{\partial t}=D\frac{\partial}{\partial p}p^3\frac{\partial }{\partial p}\left(\frac{f}{p} \right)+D\frac{\partial}{\partial \mu}\left(1-\mu^2\right)\frac{\partial }{\partial \mu}f.\quad\quad
\end{eqnarray}
Here, $\mu=\cos\theta$ is the cosine of the angle between the particle momentum and the magnetic field. 

We now assume that a particle gets accelerated locally when it can be trapped by a turbulent structure. Assuming that the typical large-scale variation of the magnetic-field {\it strength} is $\Delta B=B_{max}-B_{min}$, we estimate the trapping angles as $\mu^2<\mu_0^2= \Delta B/B_{max}$. The particles ``leak'' from the acceleration region due to pitch-angle scattering when their pitch angle cosines exceed $\mu_0$. 
In order to find the steady-state solution of Eq.~(\ref{fp_2}), we calculate the lowest eigenvalue of the pitch-angle diffusion operator in Eq.~(\ref{fokker_planck}),
\begin{eqnarray}
\label{eigen_value_problem}
-\frac{\partial}{\partial \mu}\left(1-\mu^2\right)\frac{\partial }{\partial \mu}f=\lambda f. 
\end{eqnarray}
In general, it should be supplemented by a boundary condition of the form $\left(f+a\, \partial f/\partial \mu\right)_{|\mu|=\mu_0}=0$ ensuring that the distribution function matches at the boundary the distribution function of non-accelerated particles.\footnote{For $a\to 0$ this transforms into the Dirichlet boundary condition, while for $a\to \infty$ into the Neumann boundary condition.} As the contrast between the accelerated and non-accelerated particles is expected to increase at larger $\gamma$, the parameter $a$ may in principle vary (slowly decline) with $\gamma$. For our simplified treatment, we choose  $f_{|\mu|=\mu_0}=0$, which we expect to be valid asymptotically at large $\gamma$. The eigenvalue can then be found perturbatively in the small parameter~$\mu_0^2$, which gives, up to the first order: 
\begin{eqnarray}
\lambda=\frac{\pi^2}{4 \mu_0^2}-\frac{1}{2}-\frac{\pi^2}{12}.   
\end{eqnarray}
The momentum diffusion equation now takes the form
\begin{eqnarray}
\frac{\partial f(p)}{\partial t}=D\frac{\partial}{\partial p}p^3\frac{\partial }{\partial p}\left(\frac{f}{p} \right)-D \lambda f. 
\label{energy_diffusion}
\end{eqnarray}
The last term in the right-hand side describes the loss of particles from the acceleration region of the phase space, while the first term describes the particles supply into this region and their acceleration (we remind that Eqs.~(\ref{fokker_planck}) and (\ref{energy_diffusion}) are valid only in the ultrarelativistic sector of energies, $p\approx \gamma mc$ and $\gamma \gg 1$). The steady-state solution of Eq.~(\ref{energy_diffusion}) can then be found as $f(\gamma)d\gamma\propto \gamma^{-\alpha}d\gamma$, where
\begin{eqnarray}
\alpha=\sqrt{\lambda+1}=\sqrt{\frac{\pi^2}{4} \frac{B_{max}}{\Delta B}+\frac{1}{2}-\frac{\pi^2}{12}}.
\label{alpha_turb}
\end{eqnarray}

The particle distribution function is, therefore, non-universal in our model, in that it depends on the magnetic compressibility of turbulence, $\kappa\equiv \Delta B/B_{max}$. 



\section{Numerical results} We conduct 2D simulations of a pair plasma with the fully relativistic particle-in-cell code VPIC \cite[][]{bowers2008}. The magnetic and electric fields as well as particle velocities, have three vector components but vary only in the $x-y$ plane. We perform decaying turbulence runs with a uniform magnetic field, ${\bf B_0}=B_0{\hat z}$, and the root-mean-square value of initial magnetic fluctuations, $\delta B_{0}=\langle\delta B^2({\bf x}, t=0)\rangle^{1/2}$.
We define the two plasma magnetization parameters as: 
\begin{eqnarray}
\sigma_0=\frac{B_0^2}{4\pi n_0w_0mc^2} \quad \mbox{and}\quad
{\tilde \sigma}_0=\frac{(\delta B_0)^2}{4\pi n_0w_0mc^2},
\end{eqnarray}
where $n_0$ is the initial uniform density of {\em each} species, 
$w_0=K_3(1/\theta_0)/K_2(1/\theta_0)$, where 
$K_\nu(z)$ is the modified Bessel function of the second kind, and $\theta_0=kT_0/mc^2$ is the normalized initial temperature.\footnote{In the ultrarelativistic limit $\theta_0\gg 1$, we get $w_0\sim 4\theta_0$, while in the non-relativistic case, $\theta_0\ll 1$, we have $w_0\sim 1$.} The plasma is initialized with an isotropic Maxwell-J\"uttner distribution with $\theta_0=0.3$. Our runs are summarized in Table~{\ref{table}}. 

\begin{table}[h!]
\centering
\vskip5mm
\hskip-2.0cm \begin{tabular}{c c c c c c c} 
\hline
Run & $\sigma_0$ & ${{\tilde \sigma}_0 }$ & $\left({B_0}/{\delta B_0}\right)^2$ &  $\omega_{pe}\delta t$ & $\alpha$  & $\alpha $    \\
 &  & & & & {\footnotesize (pred.)} & {\footnotesize (meas.)} \\ 
\hline
I & 0.63 & 10 & 1/16 & 0.04 & 2.0 & 2.7 \\ 
II & 2.5 & 40 & 1/16 & 0.04 & 2.0 & 2.1 \\ 
III & 10 & 10 & 1 & 0.04 & 2.9 & 3.0\\
IV & 40 & 40 & 1 & 0.02 & 2.8 & 2.7 \\
V & 90 & 10 & 9 & 0.02 & 5.1 & -- \\
VI & 360 & 40 & 9 & 0.02 & 5.2 & 4.8 \\
\hline
\end{tabular}
\caption{Parameters of the runs and the corresponding predicted and measured exponents of the particle distribution function, $f(\gamma)d\gamma\propto \gamma^{-\alpha}d\gamma$.}
\label{table}
\end{table}

The simulation domain is a double periodic square with sides $L_x=L_y\approx 2010 \,d_e$, where $d_e$ is the non-relativistic inertial scale. The resolution is $N_x=N_y=16640$. The time steps are normalized to $1/\omega_{{pe}}$, where  
$\omega_{{pe}}$ is the {non-relativistic} electron plasma frequency.  All simulations have $100$ particles per cell per species. The turbulence is seeded by randomly phased magnetic perturbations of the type
\begin{eqnarray}
\delta\mathbf{B}(\mathbf{x})=\sum_{\mathbf{k}}\delta B_\mathbf{k}\hat{\xi}_\mathbf{k}\cos(\mathbf{k}\cdot\mathbf{x}+\phi_\mathbf{k}),
\end{eqnarray}
with the wave numbers $\mathbf{k}=\{2\pi n_x/L_y,2\pi n_y/L_y\}$, $n_x,n_y=1,...,8$ and $\hat{\xi}_\mathbf{k}=\mathbf{k}\times\mathbf{B}_0/|\mathbf{k}\times\mathbf{B}_0|$. All modes $\delta B_{\mathbf{k}}$ have same amplitudes. 
The crossing time of the simulation box is then approximately 8 large-scale dynamical times, {$l/c$, where $c$ is the speed of light and $l=2\pi/k_{x,y}\,(n=8)=L_{x,y}/8$}. By this time, quasi-steady states for the distributions of fields and particles establish in all the runs except in the strong field case that requires longer running time (as explained below). 

\begin{figure}[htb!]
\includegraphics[width=\columnwidth,height=0.7\columnwidth]{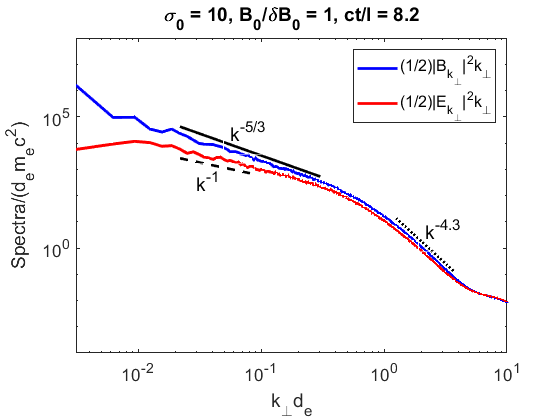}
\includegraphics[width=\columnwidth,height=0.7\columnwidth]{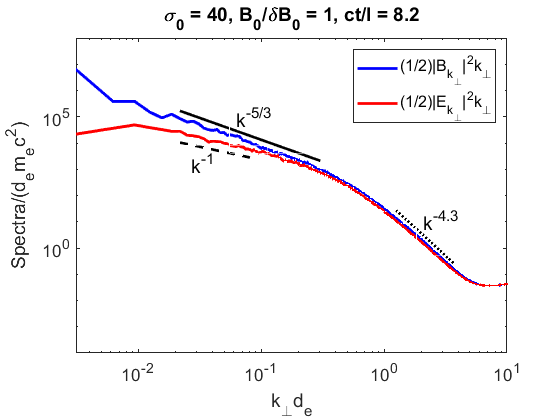}
\caption{The magnetic and electric spectra in the 2D runs with different magnetizations (runs III \& IV). 
The spectrum of the electric field approaches that of the magnetic field at small scales as the plasma magnetization increases. 
The lines indicating power laws are given for reference. }
\label{EB}
\end{figure}

We observe that in spite of different levels of initial magnetic perturbations, the resulting bulk turbulent fluctuations have comparable and rather moderate gamma factors. For example, $\langle{\tilde \gamma} \rangle\approx 1.07$ in the $\sigma_0=10$ run, and $\langle {\tilde \gamma} \rangle \approx 1.08$ in the $\sigma_0=40$ run, while the resulting particle kinetic energy is significantly larger in the second case, {with $\langle \gamma \rangle\approx2.9$ in the $\sigma_0=10$ run and $\langle \gamma \rangle\approx10$ in the $\sigma_0=40$ run.} This is consistent with our phenomenological discussion in previous sections. The spectra of magnetic fluctuations are close to that of non-relativistic Alfv\'enic turbulence, see Figure~\ref{EB}. This figure also shows that the stronger the magnetization, the more closely the spectrum of the electric field follows that of the magnetic fluctuations at small scales, also in agreement with our qualitative picture. The flatter spectrum of the electric field at large scales may be explained by the fact that in a good conductor, the electric field fluctuations are proportional to the fluctuations of the velocity field. At large scales, the velocity fluctuations are mildly relativistic, so they are limited by $c$. In this limit, the second-order velocity structure function scales as  $S_2(l)=\left\langle({\bf v}({\bf x}+{\bf l})-{\bf v}({\bf x}) \right)^2\rangle\sim c^2=\mbox{const}$, and its Fourier transform (the velocity spectrum) has the scaling~$k^{-1}$. 

\begin{figure}[htb!]
\includegraphics[width=\columnwidth,height=0.7\columnwidth]{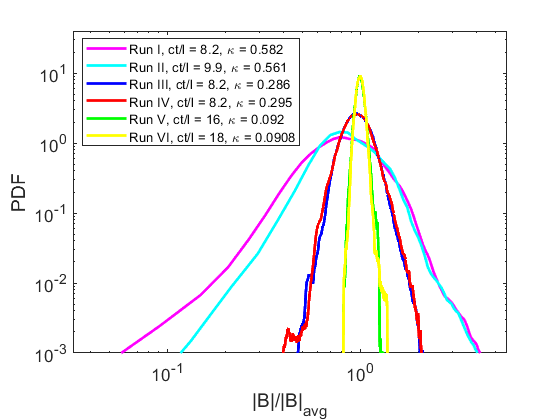}
\caption{The magnetic-field strength distribution functions and the corresponding magnetic compressibilities $\kappa$. 
The distribution is independent of the magnetization parameter, but depends on the relative strength of the applied magnetic field. 
}
\label{B_dist}
\end{figure}

The magnetic field strength distribution functions and corresponding turbulent magnetic compressibilities are presented in Figure~\ref{B_dist}. We calculate the magnetic compressibility, $\kappa=\Delta B/B_{max}$, by evaluating the mean and standard deviation of magnetic strength distributions in the simulation box, so that $B_{min}={\langle B\rangle }-(\delta B)_{rms}$, $B_{max}={\langle B\rangle }+(\delta B)_{rms}$, and $\Delta B=2(\delta B)_{rms}$.  The magnetic compressibilities are very similar in different runs with the same $\delta B_0/B_0$, they however depend on $\delta B_0/B_0$.  
\begin{figure}[htb!]
\includegraphics[width=\columnwidth,height=0.7\columnwidth]{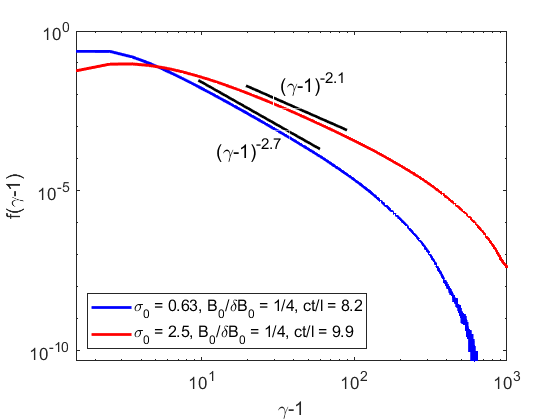}
\includegraphics[width=\columnwidth,height=0.7\columnwidth]{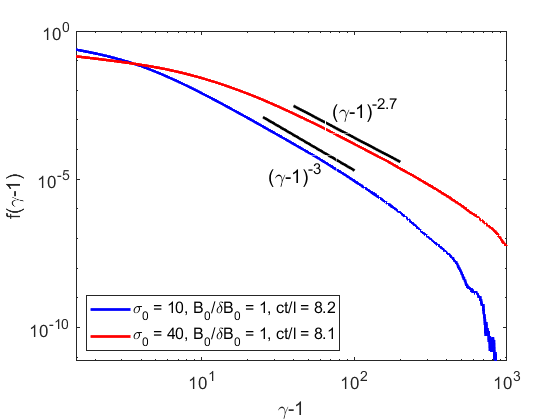}
\includegraphics[width=\columnwidth,height=0.7\columnwidth]{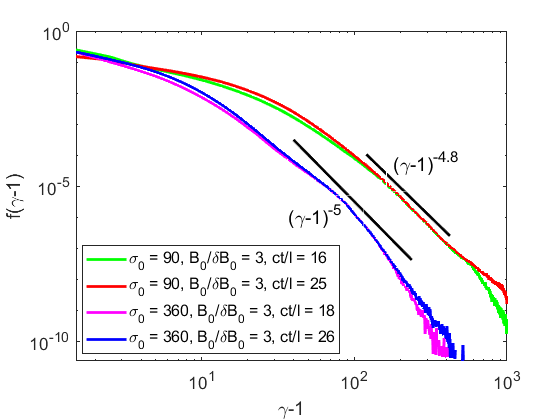}
\caption{The particle energy distribution functions in runs I, II (top), III, IV (middle), and V, VI (bottom panel). The lines indicating power-law slopes are given for reference. 
}
\label{distribution}
\end{figure}
Figure~\ref{distribution} shows the corresponding particle energy distribution functions. 
The energetic tails of the distribution functions exhibit approximate power-law behavior. As seen from Table~\ref{table}, the power-law exponents are in very good agreement with the prediction given by  Eq.~(\ref{alpha_turb}), especially in the cases of strong magnetization ${\tilde \sigma}_0=40$. {These exponents also agree well with the exponents reported by \cite[][]{comisso2018,comisso2019} for even stronger magnetization, corresponding to ${\tilde \sigma}_0=80$ in our notation.\footnote{Note that our definition of fluctuation magnetization ${\tilde \sigma}_0$ is a factor of two different from the corresponding definition in \cite[][]{comisso2018,comisso2019}. For instance, our ${\tilde \sigma}_0=40$ corresponds to $\sigma_0=20$ in Comisso \& Sironi simulations.} This means that our simulations have reached the universal asymptotic regime of strong magnetization.} In the regime of weak magnetization,\footnote{This case belongs to the asymptotic regime of weak magnetization, as it agrees with the cases of even smaller magnetization studied in \cite{zhdankin2018c}.} ${\tilde \sigma}_0=10$, our model 
underestimates the power-law exponents in the limit of weak guide field, $(B_0/\delta B_0)^2\ll 1$. This may indicate that the assumptions of the model, such as the isotropy of the random noise in Eq.~(\ref{eta}), the relation of the global magnetic compressibility $\kappa$ to the local trapping angles $\mu_0$,  and simplified boundary conditions in Eq.~(\ref{eigen_value_problem}), need to be modified for the low  magnetization, weak guide-field cases.   

The $\sigma_0=90$ and $\sigma_0=360$ runs are still evolving at $t=8\,l/c$; they start approaching quasi-steady states only at about $t=16\,l/c$.      
The absence of efficient particle acceleration (as well as well defined power-law tails) may possibly be explained by the relatively low magnetic compressibility and a smaller number of spontaneously created trapping regions where particles are accelerated. 
We also note that the plasma gets heated due to conversion of the initial energy of magnetic fluctuations into kinetic energy of particles.  Equating the initial energy of magnetic fluctuations $(\delta B_0)^2/8\pi$ to the resulting thermal kinetic energy of particles, $2n_0mc^2\gamma_{th}$, we estimate $\gamma_{th}\approx (w_0/4){\tilde \sigma}_0$, which gives for the typical gyroradius and inertial scale: $\rho_e/d_e\approx (1/\sqrt{2})\delta B_0/B_0$.  For a particle gyroradius to exceed $d_e$, we therefore need to require $\gamma >(\sqrt{2}B_0/\delta B_0)\,\gamma_{th}/\sin\theta$, where $\theta$ is the particle pitch angle (we can use, for example, the rms value $(\sin\theta)_{rms}= \sqrt{2/3}$). This estimate agrees with the slight breaks in the non-thermal tails in the bottom panel of Fig.~\ref{distribution}. Therefore, for a stronger guide field there are fewer particles whose gyroradii belong to the inertial interval of turbulence, which may lead to a less efficient acceleration.

\section{Conclusions} 
 Relativistic turbulence has been recently studied as a mechanism of particle acceleration that is alternative or complementary to previously considered particle energization by shocks~\cite[e.g.,][]{marcowith2016} or magnetic reconnection~\cite[e.g.,][]{guo2020,drake2013}. {In fact, reconnection is an inherent part of magnetic plasma turbulence   \cite[e.g.,][]{loureiro2017a,loureiro2018,mallet2017a,mallet2017,vech2018,comisso2018a,walker2018,dong2018,boldyrev2019,vega2020} and it plays  a significant role at the initial stages of particle acceleration \cite[e.g.,][]{comisso2019}.} In the present  work, we have argued that magnetically dominated relativistic turbulence quickly evolves into the state of mildly relativistic turbulence of relativistically hot plasma. We have proposed that non-thermal particle energization in such turbulence is governed by the so-called magnetic compressibility, which is related to the ability of a random turbulent flow to create magnetic structures that can trap and scatter particles. The statistics of such scattering structures are not currently well understood. We, however,  estimate the characteristic local trapping angles based on numerical measurements of global magnetic-field strength distribution. Our analytical model predicts the power-law particle distribution functions that agree well with the results of numerical simulations of magnetically dominated turbulence. Cases with weak magnetization show deviations from the predictions of the model, indicating that in this case, additional factors must be taken into account.
 
 
 Relativistic plasma turbulence is present in some astrophysical sources, so one can expect that the results of this work may explain particle acceleration in them. Intriguingly, analysis of the X-ray synchrotron emission in the Crab nebula reveals an electron energy distribution with a running power-law index going from $\alpha \sim 2.4 - 2.5$ at the energies $\gamma \sim 10^4$ to $\alpha \sim 3 - 3.2$ at the energies $\gamma \gtrsim 10^6$ \cite[][]{atoyan1996}, which may be consistent with our results.

\paragraph{Acknowledgment}
The work of CV and SB was partly supported by NSF under Grants PHY-1707272 and PHY-2010098, by NASA under Grant NASA 80NSSC18K0646, and by the Wisconsin Plasma Physics Laboratory (US Department of Energy Grant DE-SC0018266). VR was partially supported by NSF/DOE Partnership in Basic Plasma Science and Engineering through grant DE-SC0019315. MM acknowledges the NSF grant No. PHY-2010109 and the DOE EPSCOR grant No. DE-SC0019474. Computational resources were provided by the Texas Advanced Computing  Center at the University of Texas at Austin (XSEDE Allocations No. TG-PHY110016 and TG-ATM180015).

\bibliography{references}

\end{document}